# Relaxor ferromagnetic behavior below the antiferromagnetic transition in $La_{0.5}Ca_{0.5}MnO_3$

R. Mahendiran[*], A. Maignan, C. Martin, M. Hervieu, and B. Raveau

Laboratoire CRISMAT, ISMRA, Université de Caen, 6 Boulevard du Maréchal Juin,

Caen Cedex, Caen-14050, France

A detailed reinvestigation of magnetization in $La_{0.5}Ca_{0.5}MnO_3$ reveals that although a field of H = 7 T applied at T = 5 K after zero field cooling is insufficient to convert the low temperature charge ordered antiferromagnetic phase into ferromagnetic, annealing in a magnetic field as small as H = 1 T induces ferromagnetic clusters in the charge ordered matrix. The volume phase fraction of the ferromagnetic clusters increases nearly linearly with the annealing field until $H_{an}$ = 4 T and then changes dramatically from ≈ 11 % at 5 T to ≈ 60 % at $H_{an}$ = 7 T. This is analogous to the field-induced micro to macro polar domains in relaxor ferroelectrics. It is suggested that charge ordered clusters coexist with ferromagnetic phase above the Neel temperature ($T_N$) and these clusters which transform into ferromagnetic by external magnetic fields are supercooled below $T_N$. Our results will be relevant to the observation of many orders of decrease in the resistivity found for H << $H_C$, where $H_C$ is the critical field for metamagnetic transition.

[*] Present address: 104 Davey Laboratory PMB 054, Department of Physics, Pennsylvania State University, University Park, PA-16802. E-mail: mur5@psu.edu

$La_{0.5}Ca_{0.5}MnO_3$ is one of the first perovskite manganates discovered to undergo a series of magnetic and electronic transitions with lowering temperature from a charge disordered paramagnetic to ferromagnetic state ($T_C$ = 230 K) followed by a charge – orbital ordered antiferromagnetic insulating state below $T_{CO} = T_N$ = 150 K.[1] There has been revival of interest in this compound following the discovery of the collapse of the insulating state in an external magnetic field.[2] The insulating antiferromagnetic phase transforms into a ferromagnetic metallic phase in a sufficiently high magnetic field ($H > H_C$, where $H_C$ = 10-15 T is the critical field for metamagnetic transition) resulting in several orders of magnitude decrease in resistivity. It is widely believed that the charge-orbital ordered state also melts simultaneously with the collapse of the antiferromagnetic state although it is still not clear which drives what. However, a large value of magnetic field (H > 10 T) needed to destroy the charge ordering poses a severe problem for its technological exploitations. Is there any way to induce a large magnetoresistance even if $H << H_C$ ? A recent study by Parisi et al.[3] and an earlier work[4] by us indeed show a possibility of obtaining a large magnetoresistance in a field of H = 1 T ($<< H_C$) at low temperatures. It was found that field cooling enhanced magnetoresistance than when zero field cooled. In our earlier investigation[4], the resistivity and magnetization were measured on two different pieces of samples from two different batches making it difficult to correlate the properties. In this letter, we reinvestigate $La_{0.5}Ca_{0.5}MnO_3$ compound carefully by measuring the physical properties on a single piece of the sample. It is shown that $La_{0.5}Ca_{0.5}MnO_3$ is a magnetic analogue of relaxor ferroelectrics in which nanometer size ferroelectric domains in zero field become macroscopic in size when the sample is cooled in an external electric field.[5,6]

We measured four probe resistivity and magnetization of $La_{0.5}Ca_{0.5}MnO_3$ with a Quantum Design physical property measuring system and SQUID magnetometer. Measurements were taken while warming in a field after cooling the sample to the lowest temperature in zero field (ZFC mode) before establishing the field at T = 5 K and during cooling in a field from a high temperature (FCC) as well as during warming after field cooling (FCW). Isothermal magnetization at 5 K was recorded in ZFC mode as well as cooling under different magnetic fields.

Fig. 1(a) shows the temperature dependence of the magnetization in $La_{0.5}Ca_{0.5}MnO_3$ in ZFC mode (closed symbols) and in the field cooled mode during cooling (FCC) and warming (FCW) as indicated by arrows. When the sample is field cooled in H = 2 T, the sample undergoes a broad paramagnetic to ferromagnetic transition around $T_C$ = 230 K (determined by plotting H/M versus T) and then to an antiferromagnetic state. The latter transition is characterized by the onset of a rapid drop in the FCC M(T) at $T_N$ = 150 K. Upon warming from 5 K, the FCW M(T) exhibits a hysteresis with a width of ≈ 40 K centered around 135 K which suggests that the magnetic transition is first order. An irreversibility between the ZFC and FCW magnetizations occur around $T_N$ below which the FCW M(T) curve lies above the ZFC curve. Such an irreversibility between the ZFC and FCW magnetizations persists even at H = 7 T and surprisingly the difference between them increases with H ($\Delta M = M_{FCW} - M_{ZFC}$ = 0.39 and 1.88 $\mu_B$ at T = 5 K for H = 2 and 7 T, respectively). This is opposite to what is generally observed in a cluster or a spin glass in which differences narrow with increasing magnetic field values. The remarkable behavior of $La_{0.5}Ca_{0.5}MnO_3$ is due to the field-induced growth of ferromagnetic clusters as will be discussed soon.

We show the temperature dependence of the resistivity in zero external field in the inset of Fig. 2(a). The $\rho(T, H = 0\ T)$ increases with lowering temperature but does not show metallic behavior below the Curie temperature and increases abruptly at the Neel temperature ($T_N = 150$ K) in agreement with earlier reports.[2] The $\rho(T, H = 0\ T)$ below 40 K exceeds our instrumental limit and hence not shown. The main panel of Fig. 2(a) and Fig. 2(b) shows $\rho(T, H)$ in ZFC as well as FCC and FCW modes. The zero field cooling or field cooling has negligible effect on $\rho(T, H)$ below H = 0.6 T (we made measurements for different cooling or annealing fields, $H_{an}$ = 0.1, 0.2, 0.5 and 0.6 T but data are not shown here for clarity). While the ZFC $\rho(T)$ at H = 0.75 T is identical to the $\rho(T, H = 0\ T)$ curve however, a dramatic decrease in $\rho(T)$ seems to occur below T = 40 K when the sample is field cooled. This dramatic change *can not be* due to melting of the CO phase since H in excess of 7 T is needed to melt the charge order (see the discussion later). Several orders of decrease in the low temperature resistivity are also seen when the sample is cooled in a field higher than 0.75 T but the ZFC $\rho(T)$ continues to show insulating like behavior even at the highest field (H = 7 T). For example, $\rho \approx 25$ k$\Omega$ cm at H = 7 T and T = 5 K in the ZFC mode whereas $\rho \approx 9$ m$\Omega$ cm at the same temperature when annealed in a field of 7 T.

In order to have a better understanding of the annealing field dependent resistivity behavior, we carried out magnetization isotherms at 5 K in ZFC and FC modes. Figure. 3(a) compares the FC M(H) for H ≤ 5 T with the ZFC M(H). The ZFC M(H) has a very small ferromagnetic component at low fields (H < 0.5 T)[7] and increases nearly linearly with H up to the highest field of H = 7 T where it reaches a small value of M = $0.16\mu_B$. This value is much smaller than the theoretically expected magnetic moment of $M_{sat}$ = 3.5 $\mu_B$ if the whole volume of the sample were ferromagnetic. This result clearly suggests that the sample has not undergone a metamagnetic transition (i.e., field induced AF to FM transition) in the available field range (H ≤ 7 T). However, when the sample is magnetically annealed (i.e. field cooled) at H = 1 T from T > 200 K, there is a rapid increase of M(H) at low fields (H < 0.5 T) and then M increases with H without saturation. The field cooled curve at $H_{an}$ = 1 T exhibits negligible remanence ( ≈ 0.02 $\mu_B$) and coercive field (H ≈ 10 Oe). The field cooled M(H) behavior mimics that of a superparmagnetic cluster (i.e a short range ferromagnetic cluster).[8] Interestingly, annealing the sample magnetically at a higher field enhances magnetization values over all values of the magnetic field. For example, the value of M at the maximum field increases from ≈ $0.11\mu_B$ at $H_{an}$ = 1 T to ≈ $2.13\mu_B$ at $H_{an}$ = 7 T. The M(H) curves for $H_{an}$ ≥ 5.5 T also appear to saturate at higher fields. We can expect a similar trend in M(H) curves even for H > 7 T until $M_{sat}$ = $3.5\mu_B$ is reached. We have not found increase in the remanent magnetization or coercive field with increasing $H_{an}$. These remarkable behaviors of the FC M(H) data suggest that density and possibly sizes of FM clusters increase with increasing $H_{an}$. We can estimate the spontaneous magnetic moment ($M_0$) of these FM clusters by extrapolating the linear part of the high field M versus $H^{-1}$ to H =

∞ , and the volume phase fraction of the FM phase from $M_{sat} = X_{FM}M_{FM}$ where $M_{sat}$ = 3.5µB and $X_{FM}$ is the phase fraction of the FM clusters. Here, we assumed that the contribution of the magnetization from the charge ordered antiferromagnetic is negligible relative to that of ferromagnetic clusters. The inset of Fig. 3(b) shows the evolution of $X_{SP}$ with $H_{an}$. The $X_{FM}$ initially increases linearly with $H_{an}$ until 4 T where it reaches only 7 %. A more dramatic increase occurs between $H_{an}$ = 5 and 7 T where it changes from ≈ 11 % to ≈ 60 %. In accordance with the magnetization, the isothermal magnetoresistance at 5 K also shows dramatic effects as shown in Fig. 4. As the field is increased from 0 T, the ZFC ρ(H) is not measurable below H = 4 T and in the field range 4 – 6 T it decreases gradually and then more abruptly as H is approached 7 T. While decreasing H, ρ(H) increases a little but remains much lower than the starting value. The high resistive virgin state is not recovered even when the field is reversed. However, annealing in a field of H = 0.75 T or a higher field leads to dramatic decrease in the resistivity compared to the ZFC mode as can be clearly seen in Fig. 4.

To elucidate the origin of the observed field cooling effect, we have to understand what happen when the sample is subjected to a high magnetic field at $T > T_N$. Fig. 5 shows M(H) at two temperatures for $T_N \leq T \leq T_C$. The M(H) curve at T = 175 K shows a rapid increase at very low fields (H < 10 mT) followed by a gradual increase until H = 1.3 T due to alignment of ferromagnetic domains. However, a field-induced metamagnetic transition characterized by a rapid increase in the magnetization occurs between 1.3 and 6 T. This field-induced metamagnetic transition is first order as suggested by its hysteretic behavior (seen also in T = 185 K data). A possible origin of this field-induced metamagnetic transition above $T_N$ is that charge ordered clusters of various sizes are already imbedded in the ferromagnetic phase. As the field increases above a certain value, charge ordering melts within these clusters and these clusters become ferromagnetic since the delocalized $e_g$-charge carriers promote ferromagnetic correlations among the localized $t^3_{2g}$ spins. When the temperature is lowered in presence of a field, these field-induced FM clusters created at higher temperatures do not convert back into AF while the rest of the matrix transforms into antiferromagnetic and charge ordered phase. This means, the field-induced FM clusters at high temperatures are supercooled below its melting temperature (= $T_N$). The number density of these field-induced FM clusters initially increases with increasing $H_{an}$ but at higher $H_{an}$ (> 5.5 T) these clusters coalesce and form bigger clusters or macroscopic domains. This behavior is similar to what is observed in relaxor ferroelectrics where the zero field cooled state is made up of nanometer size polar regions which grow into macrodomains and shows normal ferroelectric behavior upon field cooling.[5] Hence, $La_{0.5}Ca_{0.5}MnO_3$ can be considered as a magnetic analogue of the relaxor ferroelectric. Kimura et al.[6] also observed similar magnetic behavior in $Nd_{0.5}Ca_{0.5}Mn_{0.98}Cr_{0.02}O_3$ and attributed the relaxor behavior due to random electric fields created by the lack of $e_g$-orbital of $Cr^{3+}$:$t^3_{2g}$ and Li et al.[9] reported in $Nd_{0.5}Ca_{0.5}Mn_{0.98}Ga_{0.02}O_3$. However, our results suggest Mn-site doping is not essential for the observation of the relaxor ferromagnetic like behavior.

A comparison of the magnetization and resistivity data suggests an unusual feature: There is more than four orders magnitude decrease in the resistivity relative to the zero field cooled value when $H_{an}$ = 4 T although the ferromagnetic phase fraction ($X_{FM}$ = 7 % at 4 T) is much below the limit of percolation threshold ($X_P \approx 16$ %) in three dimensions.[10] In particular, the dramatic decrease in the resistivity found at low temperatures when $H_{an}$ = 0.75 T relative to the zero field value is remarkable. There are two possibilities which can explain the above observation. The percolation threshold Xp = 16 % is valid only if the metallic and insulating regions are assumed to be spherical and their radius are equal (RM = RI, where RM and RI are radii of metallic and insulating spheres). However if $R_M \ll R_I$, or they are elongated instead of spherical, Xp can decreases to as low as 0.1 %.[11] It is possible that the FM clusters are not spherical in size but in elongated or in filamentary form and they percolate and run through the volume of the sample. Mayr et al.[12] shown that a small change in the volume fraction of metallic phase in a model consisting of parallel network of metallic and insulating resistors can have dramatic decrease in the resistivity. Another possibility is that the FM clusters are isolated and the large magnetoresistance observed below the percolation limit is due to tunneling of charge carriers between FM clusters through the intervening charge ordered regions.

In conclusion, we have shown that $LaCa_{0.5}MnO_3$ is a very interesting system in which magnetic field cooling promotes creation and enlargement of ferromagnetic clusters in the charge-ordered antiferromagnetic matrix and colossal magnetoresistance can be obtained for a field value much lower than the critical field for metamagnetic transition. It will be worth investigating whether such a field cooling effect is a common aspect in other half doped manganates.

Acknowledgements: The first author thanks MNERT (France) for financial assistance.

these clusters occupy less than 1 % of total volume in a field of 1 T after zero field cooled.

FIG. 1. Temperature dependence of the magnetization, in logarithmic scale, at different values of fields after zero field cooling (ZFC, closed symbols), during field cooling (FCC) and warming after field cooling (FCW). The directions of the temperature sweep are marked by arrows.

FIG. 2. Temperature dependence of the resistivity, $\rho(T)$ in ZFC (solid lines), and field cooled mode (closed circles) during cooling (FCC) and warming  for (a) $H \leq 3$ T and (b) $H = 5$ and 7 T. Note that several orders of decrease in the FC resistivity relative to ZFC values occurs at a field as small as $H = 0.75$ T at low temperatures. The inset of Fig. 2(a) shows $\rho(T)$ for $H = 0$ T.

FIG. 3. Field dependence of the magnetization at 5 K recorded after zero field cooling the sample to 5 K prior to the establishment of a field and after cooling the sample from T > 200 K to 5 K at different fields. (a). M(H) in ZFC mode (solid line) and field cooled modes (symbols connected by lines) for 1 T ≤ $H_{an}$ ≤ 5 T. The inset shows the percentage volume phase fraction of the ferromagnetic phase ($X_{FM}$) as a function of the annealing field ($H_{an}$). (b) M(H) isotherms for 5.5 T ≤ $H_{an}$ ≤ 7 T.

FIG. 4. Field dependence of the resistivity at T = 5 K measured after zero field cooling to 5 K (ZFC) and after annealing at different values of the magnetic field. The strength of anneling fields are mention with the pretext "F".

FIG. 5. Field dependence of the magnetization at T = 185 K and 175 K which are above $T_N$ and bleow $T_C$. Note that M(H) shows metamagnetic behavior above the Neel temperature.

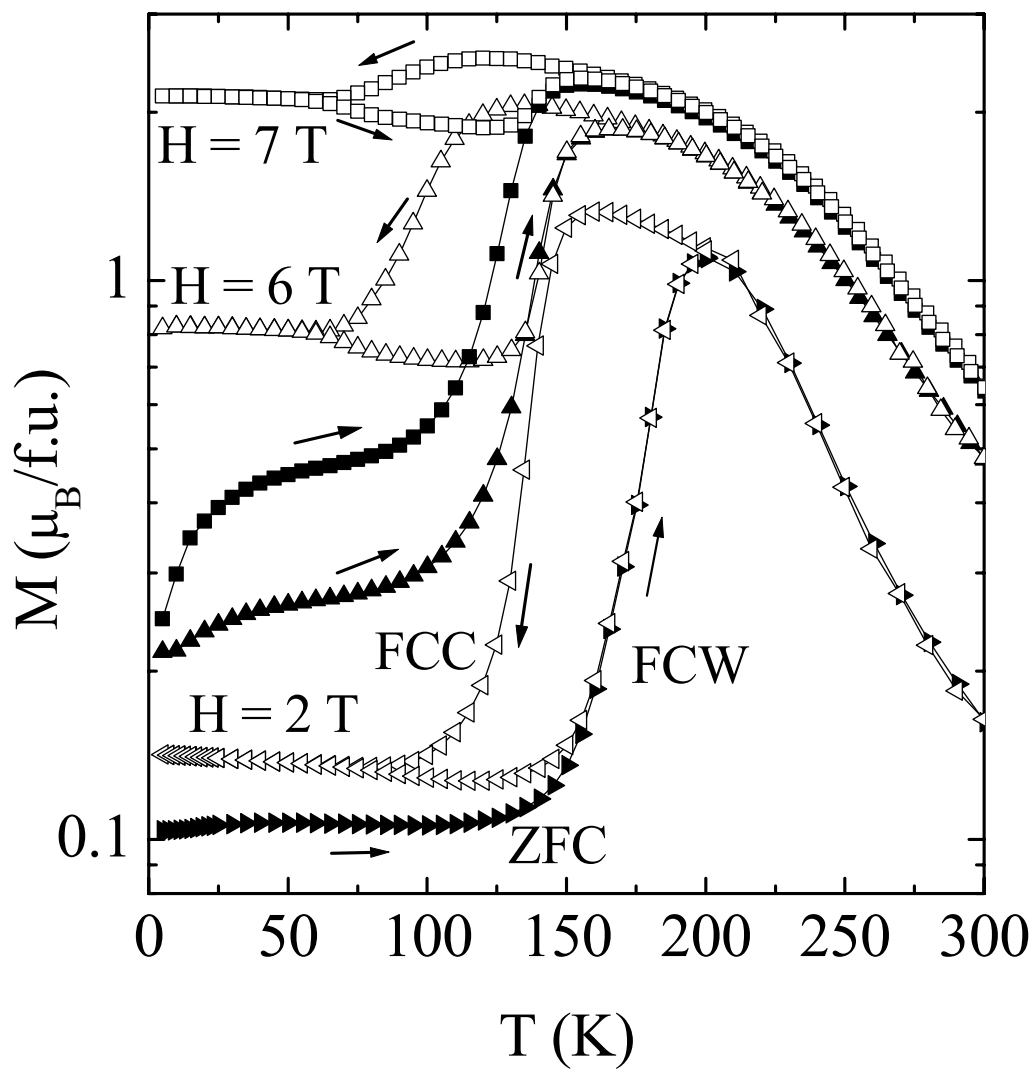

FIG. 1
R. Mahendiran et al.

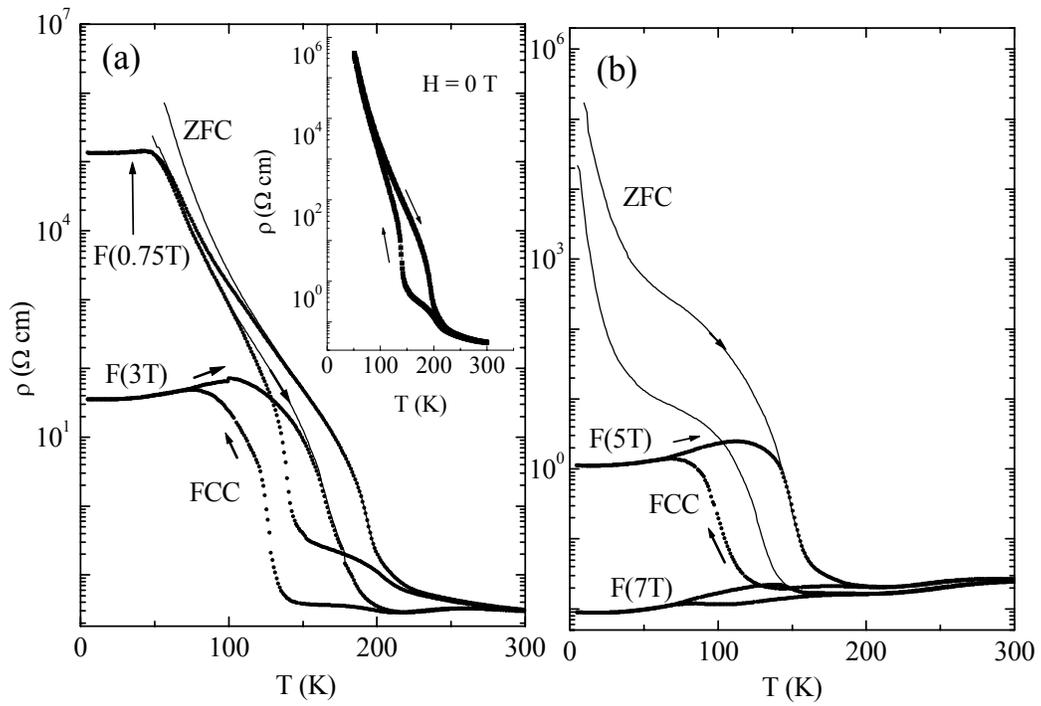

FIG. 2
R. Mahendiran et al.

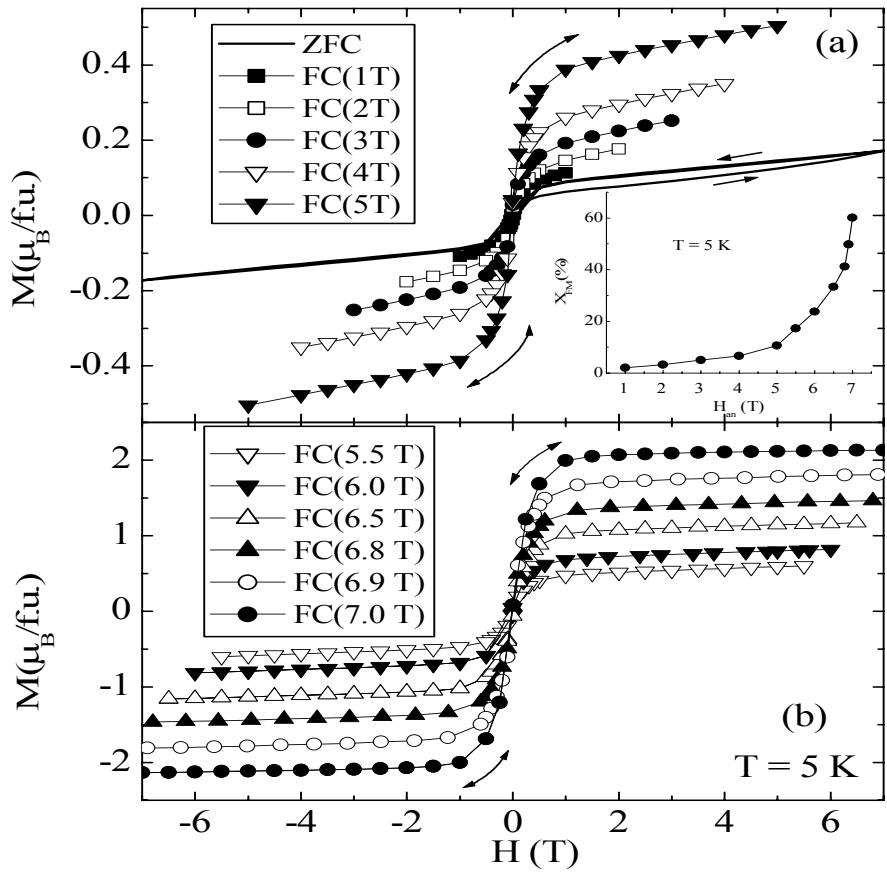

FIG. 3
R. Mahendiran et al.

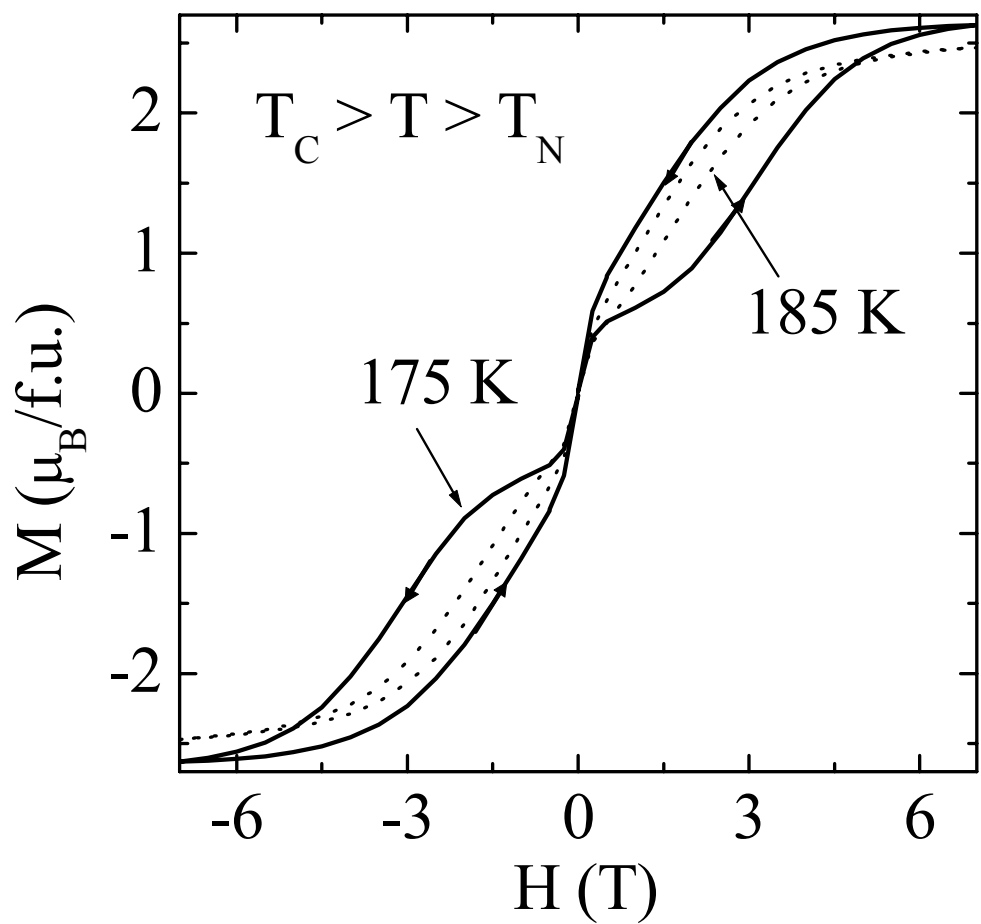

FIG. 5
R. Mahendiran et al.